\documentclass{emulateapj}

\usepackage{graphicx}	
\usepackage{amsmath}	
\usepackage{amssymb}	

\shorttitle{IMF variation in MaNGA galaxies}
\shortauthors{Hongyu~Li et al.}

\begin{document}
\title{SDSS-IV MaNGA: variation of the stellar initial mass function in spiral and early-type galaxies}

\author{\mbox{Hongyu~Li\altaffilmark{1,2}}}
\author{\mbox{Junqiang Ge\altaffilmark{1}}}
\author{\mbox{Shude~Mao\altaffilmark{3,1,4}}}
\author{\mbox{Michele Cappellari\altaffilmark{5}}}
\author{\mbox{R.~J.~Long\altaffilmark{1,4}}}
\author{\mbox{Ran~Li\altaffilmark{1,6}}}
\author{\mbox{Eric Emsellem\altaffilmark{7,8}}}
\author{\mbox{Aaron A. Dutton\altaffilmark{9}}}
\author{\mbox{Cheng Li\altaffilmark{10,3}}}
\author{\mbox{Kevin Bundy\altaffilmark{11}}}
\author{\mbox{Daniel Thomas\altaffilmark{12}}}
\author{\mbox{ Niv Drory\altaffilmark{13}}}
\author{\mbox{Alexandre Roman Lopes\altaffilmark{14}}}

\altaffiltext{1}{National Astronomical Observatories, Chinese Academy of Sciences, 20A Datun Road, Chaoyang District, Beijing 100012, China ({\tt hyli@nao.cas.cn})}
\altaffiltext{2}{University of Chinese Academy of Sciences, Beijing 100049, China}
\altaffiltext{3}{Physics Department and Tsinghua Centre for Astrophysics, Tsinghua University, Beijing 100084, China}
\altaffiltext{4}{Jodrell Bank Centre for Astrophysics, School of Physics and Astronomy, The University of Manchester, Oxford Road, Manchester M13 9PL, UK}
\altaffiltext{5}{Sub-Department of Astrophysics, Department of Physics, University of Oxford, Denys Wilkinson Building, Keble Road, Oxford, OX1 3RH, UK}
\altaffiltext{6}{Key laboratory for Computational Astrophysics, National Astronomical Observatories, Chinese Academy of Sciences, Beijing, 100012, China}

\altaffiltext{7}{Universit\'e Lyon 1, Observatoire de Lyon, Centre de Recherche Astrophysique de Lyon and Ecole Normale Sup\'erieure de Lyon, 9 avenue Charles Andr\'e, F-69230 Saint-Genis Laval, France}
\altaffiltext{8}{European Southern Observatory, Karl-Schwarzschild-Str. 2, 85748 Garching, Germany}
\altaffiltext{9}{New York University Abu Dhabi, PO Box 129188, Abu Dhabi, UAE}
\altaffiltext{10}{Shanghai Astronomical Observatory, Shanghai 200030, China}
\altaffiltext{11}{Kavli IPMU (WPI), UTIAS, The University of Tokyo, Kashiwa, Chiba 277-8583, Japan}
\altaffiltext{12}{Institute of Cosmology \& Gravitation, University of Portsmouth, Dennis Sciama Building, Portsmouth, PO1 3FX, UK}
\altaffiltext{13}{McDonald Observatory, The University of Texas at Austin, 1 University Station, Austin, TX 78712, USA}
\altaffiltext{14}{Department of Physics and Astronomy, Universidad de La Serena, Cisternas 1200, La Serena, Chile}

\begin{abstract}
We perform Jeans anisotropic modeling (JAM) on elliptical and spiral galaxies from the MaNGA DR13 sample. By 
comparing the stellar mass-to-light ratios estimated from stellar population synthesis (SPS) and from JAM, we find a similar
systematic variation of the initial mass function (IMF) as in the earlier $\rm ATLAS^{3D}$ results.  Early type galaxies (elliptical and 
lenticular) with lower velocity dispersions within one effective radius are consistent with a Chabrier-like IMF while galaxies with
higher velocity dispersions are consistent with a more bottom heavy IMF such as the Salpeter IMF. 
Spiral galaxies have similar systematic IMF variations, but with slightly different slopes and larger scatters, due 
to the uncertainties caused by higher gas fractions and extinctions for these galaxies. 
Furthermore, we examine the effects of stellar mass-to-light ratio gradients on our JAM modeling, and find that the trends from our 
results becomes stronger after considering the gradients. 
\end{abstract}

\keywords{dark matter --- galaxies: evolution --- galaxies: formation --- galaxies: kinematics and dynamics ---  galaxies: structure}



\maketitle

\section{Introduction}
Stellar mass is one of the fundamental attributes of a galaxy. Accurate estimation of stellar mass plays an important role in the 
study of a galaxy's structure, evolution and formation \citep{Cappellari2016}. Stellar population synthesis (SPS) is the most popular 
method for obtaining
the stellar mass. However, the stellar mass so obtained depends strongly on assuming a stellar initial mass
function (IMF).  Estimated stellar masses will shift on average 0.25 dex higher by changing the IMF from the Chabrier IMF  
\citep{Chabrier2003} to the Salpeter IMF \citep{Salpeter1955} (see \citealt{Panter2007,Tortora2009}). This  causes  uncertainty in
the determination of the dark matter fraction in a galaxy's central region \citep{Cappellari2006,Tortora2009}. Furthermore, whether the IMF is universal
or not has been discussed for decades \citep{Bastian2010}. The situation is becoming clearer after numerous studies based on 
line-strength indices (e.g. \citealt{Conroy2012, Spiniello2012}), strong lensing plus spatially unresolved stellar kinematics
(e.g. \citealt{Treu2010, Posacki2015}),
resolved stellar kinematics \citep{Cappellari2012, Cappellari2013a}, and the fundamental plane \citep{Dutton2013}. 
All these studies show evidence for variation of the IMF in early type (i.e. elliptical or lenticular) galaxies. 

In \citet{Cappellari2012}, the Jeans anisotropic modeling technique (JAM, \citealt{Cappellari2008}) was used, with a spatially constant stellar 
mass-to-light ratio and several different dark matter halo models, to obtain stellar mass estimates without resorting to SPS.
They used 256 early type galaxies from the $\rm ATLAS^{3D}$ integral field unit (IFU) survey, and found a systematic variation 
in the IMF with galaxy stellar mass-to-light ratio. With the increasing availability of IFUs, more and more nearby galaxies with
IFU data are becoming available, e.g. CALIFA \citep{Sanchez2012}, SAMI \citep{Bryant2015} and MaNGA \citep{Bundy2015}. 
The MaNGA DR13 \citep{DR13} sample includes 1390 galaxies of different morphologies (both early and late type galaxies), and is 
currently the largest IFU sample. Please see the following references for more details about the MaNGA instrumentation \citep{Drory2015}, 
observing strategy \citep{Law2015}, spectrophotometric calibration\citep{Smee2013,Yan2016a}, and survey execution and initial data quality 
\citep{Yan2016b}. In this paper, we take advantage of the MaNGA DR13 sample to investigate IMF variation for 
both early type and late type galaxies using a similar method to \citet{Cappellari2012}. Furthermore, we examine the effects  
of  stellar mass-to-light ratio gradients (\citealt{Portinari2010,Tortora2011}; Ge et al., in preparation) on our dynamical models 
(i.e. not using a constant stellar mass-to-light ratio). 
      
The structure of this paper is as follows. In Section~\ref{Data_sample}, we briefly introduce the galaxy sample and the modeling
methods we use. In Section~\ref{rst}, we show our results concerning the systematic variation of the IMF, and the effects of stellar 
mass-to-light ratio gradients. In Section~\ref{conclusion}, we summarize and give our conclusions.  
   
\section{MaNGA sample and Methods}\label{Data_sample}

\subsection{MaNGA sample}
We use the MaNGA Product Launch 4 (MPL4) IFU spectra
from the MaNGA DR13 SDSS-IV sample. The IFU spectra are extracted using
the official data reduction pipeline (DRP, \citealt{Law2016}) and kinematical data are extracted using the official data analysis pipeline 
(DAP, Westfall et al. in prep). From the 1390 galaxies in the MaNGA DR13 catalogue, 
we exclude merging galaxies (close galaxy pairs, extremely unsmooth structures) and galaxies with low data quality 
(with less than 20 Voronoi bins with S/N greater than 30). In total,
we are left with 816 galaxies (413 spiral galaxies; 403 elliptical and lenticular galaxies), 
more than a factor of three times larger than the 
$\rm ATLAS^{3D}$ sample. We visually select galaxies with the best data qualities and JAM modeling as our class A  subsample
(sufficient Voronoi bins, high S/N, no strong bars and spiral arms -- these galaxies will have reliable
 JAM models). There are 346 galaxies in the class A subsample. We match our whole galaxy 
sample with \textit{Galaxy Zoo~1} \citep{Lintott2008} to obtain galaxy morphologies.

\subsection{Stellar population synthesis}\label{SSP}
\begin{figure}
\includegraphics[width=\columnwidth]{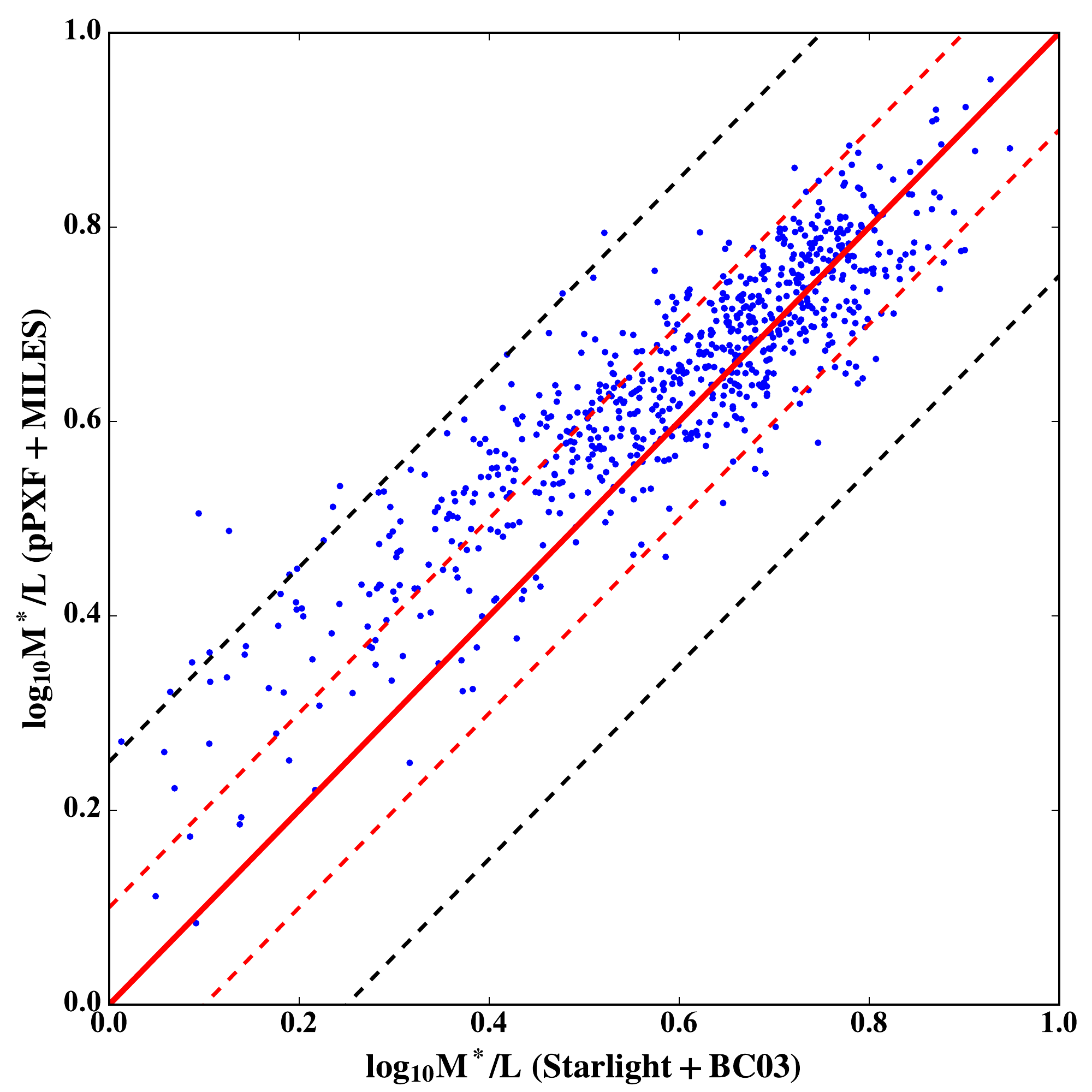}
\caption{Comparison of the stellar mass-to-light ratios for all 816 galaxies between the {\bf \scriptsize pPXF} software with MILES templates and the
 {\bf \scriptsize STARLIGHT} software with BC03
templates. The red dashed lines show a 0.1 dex difference, and the black dashed lines show a 0.25 dex difference, which is the difference between
the Salpeter and Chabrier IMFs.}
\label{cmpML}
\end{figure}

To asses the robustness of our results, we derive the stellar masses of all our galaxies using two different full spectrum fitting software and two 
different SPS template libraries. In both software packages we adopted for reference a \citet{Salpeter1955} IMF. The first stellar mass estimate
uses the {\bf \scriptsize STARLIGHT} software \citep{Fernandes2005}, in combination with the BC03 SPS templates \citep{BC03},
 while the second uses
the {\bf \scriptsize pPXF} software \citep{Cappellari2004, Cappellari2017} with the MILES-based  \citep{MILES2006} SPS models of
\citet{Vazdekis2010}. For {\bf \scriptsize STARLIGHT}, we use 25 ages (logAge = [6.00, 6.50, 6.70, 6.82, 6.94, 7.00, 7.16, 7.40, 7.60,
 7.74, 8.01, 8.21, 8.46, 8.71, 8.96, 9.11, 9.16, 9.40, 9.63, 9.80, 9.88, 10.00, 10.11, 10.18, 10.26] years), and 6 metallicities 
 ([Z/H] = [-1.7, -1.3, -0.7, -0.4, 0.0, 0.4]).  For {\bf \scriptsize pPXF}, we use 25 ages linearly-spaced in logAge (yr) between $7.8$ and
  $10.25$, and 6 metallicities  ([Z/H] = [-1.7, -1.3, -0.7, -0.4, 0.0, 0.2]).
As can be seen from Fig.~\ref{cmpML}, the stellar mass-to-light ratios obtained from different templates
and software agree well for galaxies with higher stellar mass-to-light ratios (i.e. old ages). The scatter is less than 0.1 dex and with 
nearly no bias. For galaxies with lower stellar mass-to-light ratios (i.e. young ages), {\bf \scriptsize pPXF} with MILES templates gives systematically 
higher stellar  mass-to-light ratios than {\bf \scriptsize STARLIGHT} with BC03. This is because the $\rm M^*/L$ is more degenerate for younger galaxies than 
for older ones. The reason for this is that in young galaxies the light of few bright stars can dominate the flux in a galaxy's spectrum. This makes it easy
to `hide' significant numbers of old stars, which emit a small amount of light, but contribute significantly to the mass, increasing the $\rm M^*/L$. However, the 
difference between Salpeter and Chabrier IMF is 0.25 dex, so this 0.1 dex difference
will not strongly affect our conclusions. This is shown explicitly in Section~\ref{IMF_variation}, where consistent IMF trends are presented 
using stellar masses from both STARLIGHT and pPXF. We use 0.1 dex as the uncertainty in the SPS in the following analysis.
More details about the comparison of software packages and templates can be found in Ge et al. (in preparation). 

We calculate our stellar mass-to-light ratios using Equation 2 in \citet{Cappellari2013a}
\begin{equation}
\label{ML}
\rm{(M^*/L)_{SPS}} = \frac{\sum_{j=1}^N w_j M_j^{nogas}}{\sum_{j=1}^N w_j L_j},
\end{equation}
where $\rm M_j^{nogas}$ is the stellar mass of the $j$th template,
which includes the mass in living stars and stellar remnants, but excludes the gas lost during stellar 
evolution. $\rm L_j$ is the corresponding r-band luminosity. $\rm w_j$ is the weight of the $j$th template.

Before spectrum fitting, the data cubes are Voronoi binned \citep{Cappellari2003} to a S/N=30. For all resulting Voronoi bins in each galaxy,
the two software fit for both the templates weights and for the dust extinction, adopting a \citet{Calzetti2000} reddening curve. The luminosity
 of each spatial bin is separately corrected for the measured extinction, before computing the total r-band $\rm (M*/L)_{SPS}$ for a galaxy 
 by summing the luminosity and masses of all the bins within the MaNGA field-of-view. This dust extinction corrected 
 $\rm{(M^*/L)_{SPS}}$ is directly comparable with the stellar mass-to-light ratios in JAM modeling, obtained from stellar masses
divided by the observed r-band luminosities. See Section~\ref{incAndextinct} for more discussion about the extinction and inclination effects.

\subsection{Dynamical modeling}
We perform Jeans Anisotropic Modeling (JAM, \citealt{Cappellari2008}) for all 816 selected galaxies. For a given luminosity density
(constructed by fitting a galaxy's surface brightness using {\bf mge\_fit\_sectors} software \citep{Cappellari2002}
and deprojecting it using the Multi-Gaussian Expansion (MGE) method of
\citealt{Emsellem1994}), we assume a spatially constant stellar mass-to-light ratio and a gNFW dark halo profile (also see 
\citealt{Barnab2012,Cappellari2013a}) 
\begin{equation}
        \rho_{\rm DM}(r)=\rho_s \left(\frac{r}{R_s}\right)^\gamma
            \left(\frac{1}{2}+\frac{1}{2}\frac{r}{R_s}\right)^{-\gamma-3}
\end{equation}
to construct a galaxy's total mass
model. From running JAM within an MCMC framework ({{\bf emcee}, \citealt{Foreman2013}), we find the best-fitting parameters (including
 the stellar mass-to-light
ratio $\rm (M^*/L)_{JAM}$) which give the best model matching the galaxy's observed second velocity moment map. We correct the 
cosmological surface  brightness dimming effect in our MGE by multiplying the surface  brightness by a factor $(1+z)^3$, which accounts both for 
the bolometric surface brightness dimming and the change of the band width (in the AB system). Since the MaNGA galaxies are mostly of low redshift
(median and maximum redshift of the MaNGA sample are 0.03 and 0.15 respectively), we choose not to apply K-correction (e.g. \citealt{Hogg2002}) here.
The stellar mass-to-light ratios obtained from this method are independent of SPS, and so can be used to test the variation of the IMF. The 
details of the modeling process are described in \citet{Hongyu2016}, which assesses the validity of the JAM method using 
cosmologically simulated galaxies. 
We broaden the prior for the central dark halo slope $\gamma$ from $[-1.2,0]$ to $[-1.6,0]$ to avoid the bias in the IMF, which is sensitive to 
the halo response to baryonic settling \citep{Dutton2013}.  The prior is consistent with simulated haloes in the EAGLE cosmological 
simulation \citep{Schaller2015}, as well as elliptical galaxy zoom-in simulations which find inner slopes of $\sim -1.6$ 
\citep{Dutton2015}, and spiral galaxies simulations which can have inner slopes of $\sim 0$  \citep{Tollet2016}.

\begin{figure}
\includegraphics[width=\columnwidth]{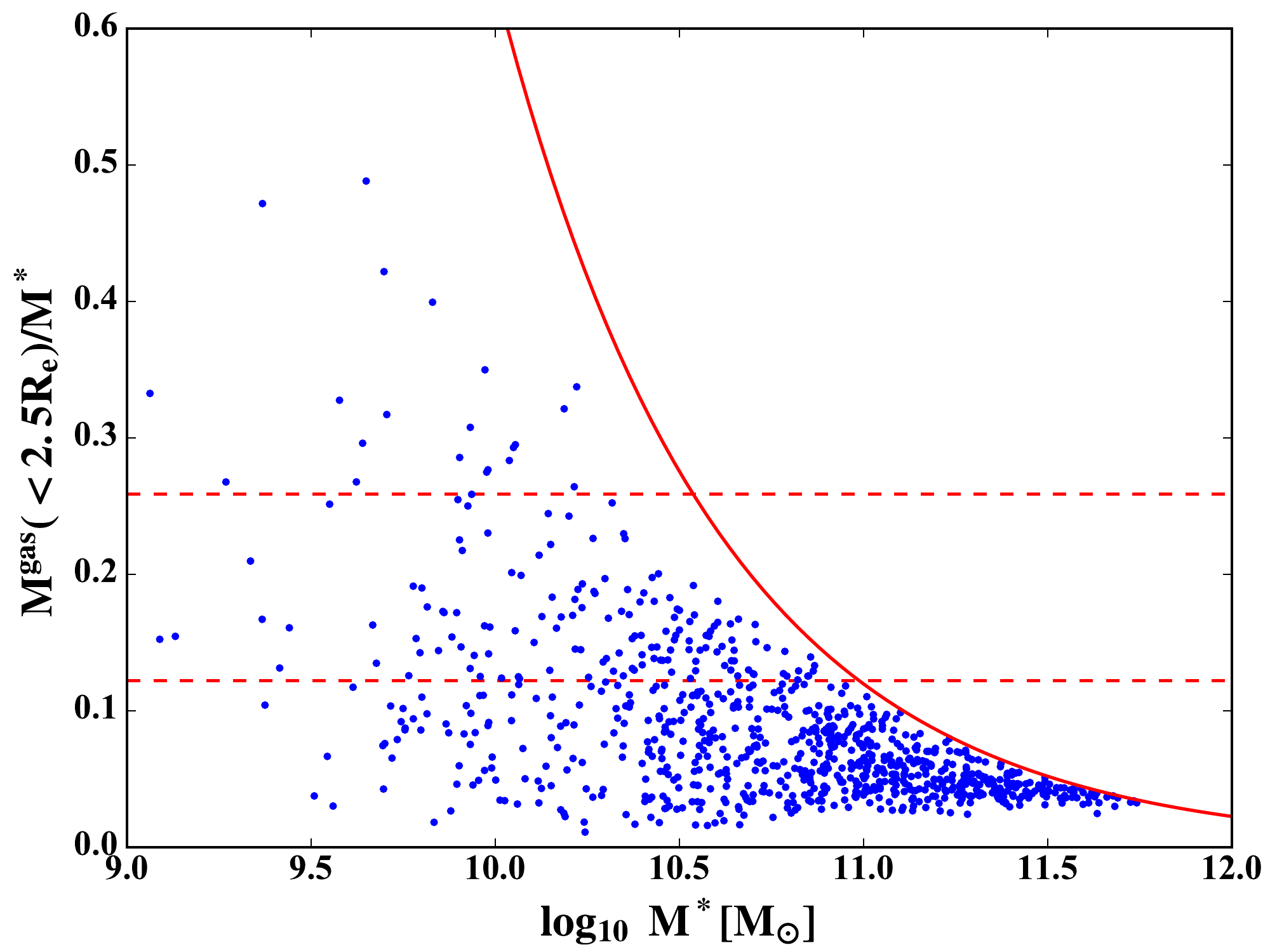}
\caption{Gas mass fraction for all the spiral galaxies. The red solid line shows the gas fraction for a given stellar mass if all the gas is 
within 2.5$\rm R_e$. The red dashed lines show the fraction at which the $\rm M^*/L_{JAM}$ will change by 0.05 dex and 0.1 dex in 
the gas correction respectively.}
\label{gas}
\end{figure}

Since spiral galaxies have higher gas fractions, especially later type spirals \citep{Combes2013,Jaskot2015,Huang2012}, we need to
consider  the gas contribution to the stellar masses derived from JAM. We perform the following steps to reduce the effects of cold gas in
spiral galaxies:
\begin{enumerate}
\item We use the $\rm M^{gas}$-$\rm M^*$ relation from (\citealt{Huang2012}, Equation 1) to estimate the gas mass for every spiral galaxy. 
The stellar masses we use in applying the Huang relationship are taken from SPS.
\item We assume the gas (HI + $\rm H_2$) mass distribution can be approximated by an exponential disk with scale length 6.1 kpc
\citep{Bigiel2012}. 
 \item We calculate the gas mass within 2.5$\rm R_e$ for every spiral galaxy using the gas mass profile described 
 above. 
 \item We use the formula below to correct for the 
effect of gas in the JAM stellar mass-to-light ratios for spiral galaxies

\begin{equation}
\rm{(M^*/L)_{JAM}^{nogas}} = \frac{\rm{M^*_{JAM}-M^{gas}(<2.5R_e)}}{L}, 
\end{equation}
where $\rm{(M^*/L)_{JAM}^{nogas}}$ is the final value that we use to investigate IMFs, $\rm M^*_{JAM}$ is the stellar mass estimated by JAM,
and $\rm M^{gas}(<2.5R_e)$ is the gas mass within $\rm 2.5R_e$. 
\end{enumerate}

In Fig.~\ref{gas}, we plot the gas mass fraction within 2.5$\rm R_e$ vs. galaxy stellar mass to show the impact of the gas correction.
As can be seen, for massive spiral galaxies
($\rm \log M^*>11.0$), the gas fraction is smaller than 10\%, which has nearly no effect on the dynamical models. As shown by 
the red dashed lines, the change of $\rm{(M^*/L)_{JAM}}$ is smaller than 0.05 dex for more than half of the galaxies, and 
smaller than 0.1 dex for more than 90\% of the galaxies.

\begin{figure*}
\includegraphics[width=\textwidth]{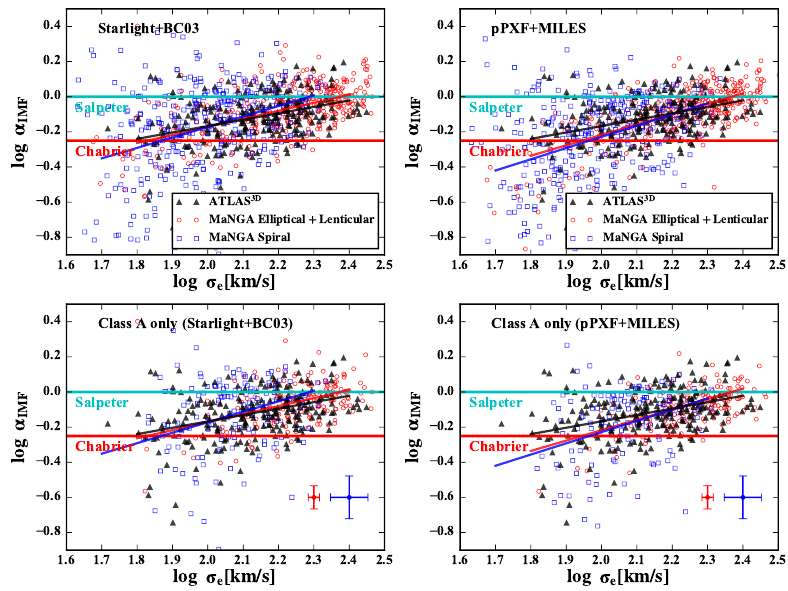}
\caption{
Systematic IMF variation for MaNGA galaxies. Upper left: the log $\alpha_{\rm IMF}$ for {\bf \scriptsize STARLIGHT} + BC03 vs. galaxy velocity dispersion.  
Lower left: class A sample only, which has the most reliable fitting among the whole sample. Upper right: the log $\alpha_{\rm IMF}$
for {\bf \scriptsize pPXF} + MILES vs. galaxy velocity dispersion. Lower right: class A sample only.  In each panel, the black triangles show the 
results from $\rm ATLAS^{3D}$,
red circles for the MaNGA elliptical galaxies and blue squares for MaNGA spiral galaxies. The solid lines show the
linear fitting results respectively. The horizontal colored solid lines show the 
positions where the stellar mass-to-light ratios from JAM equal the SPS with a Salpeter IMF (cyan) and SPS with a Chabrier IMF (red). 
The mean errors for elliptical and spiral galaxies 
are shown in the red and blue error bars
in the lower panels respectively. The error in $\rm (M^*/L)_{JAM}$ is estimated using the MCMC 1D marginalised distribution, while the error in
$\rm (M^*/L)_{SPS}$ is obtained by using different stellar templates and software in SPS. 
}
\label{IMF}
\end{figure*}

\begin{figure}
\includegraphics[width=\columnwidth]{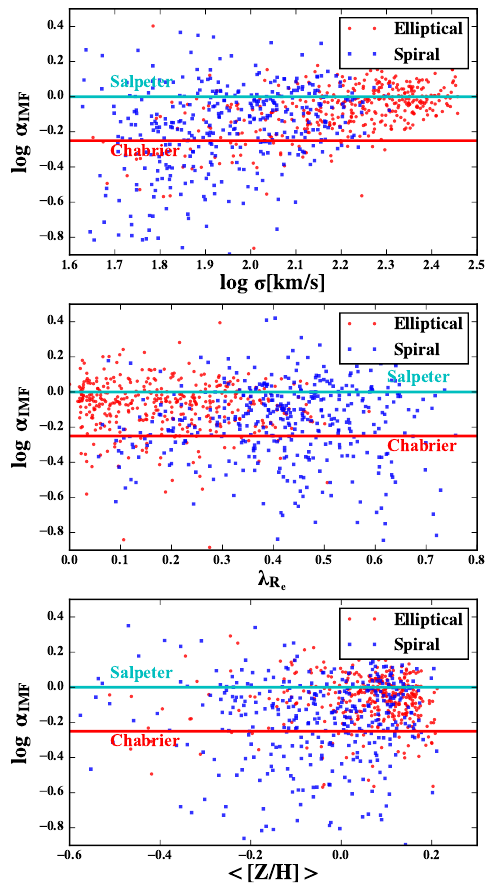}
\caption{
$\log \alpha_{\rm IMF}$ vs. $\log \sigma$ (top), $\lambda_{\rm R_e}$ (middle) and metallicity (bottom).  
Other labels and legends are the same as Fig.~\ref{IMF}.}
\label{metal}
\end{figure}

\section{Results}\label{rst}
In this section, we first show the systematic variation of the IMF and make a comparison with $\rm ATLAS^{3D}$ results. 
We then show the results from including the stellar mass-to-light ratio gradients.   

\subsection{Systematic variation of IMF}\label{IMF_variation}

\begin{figure*}
\begin{center}
\includegraphics[width=0.8\textwidth]{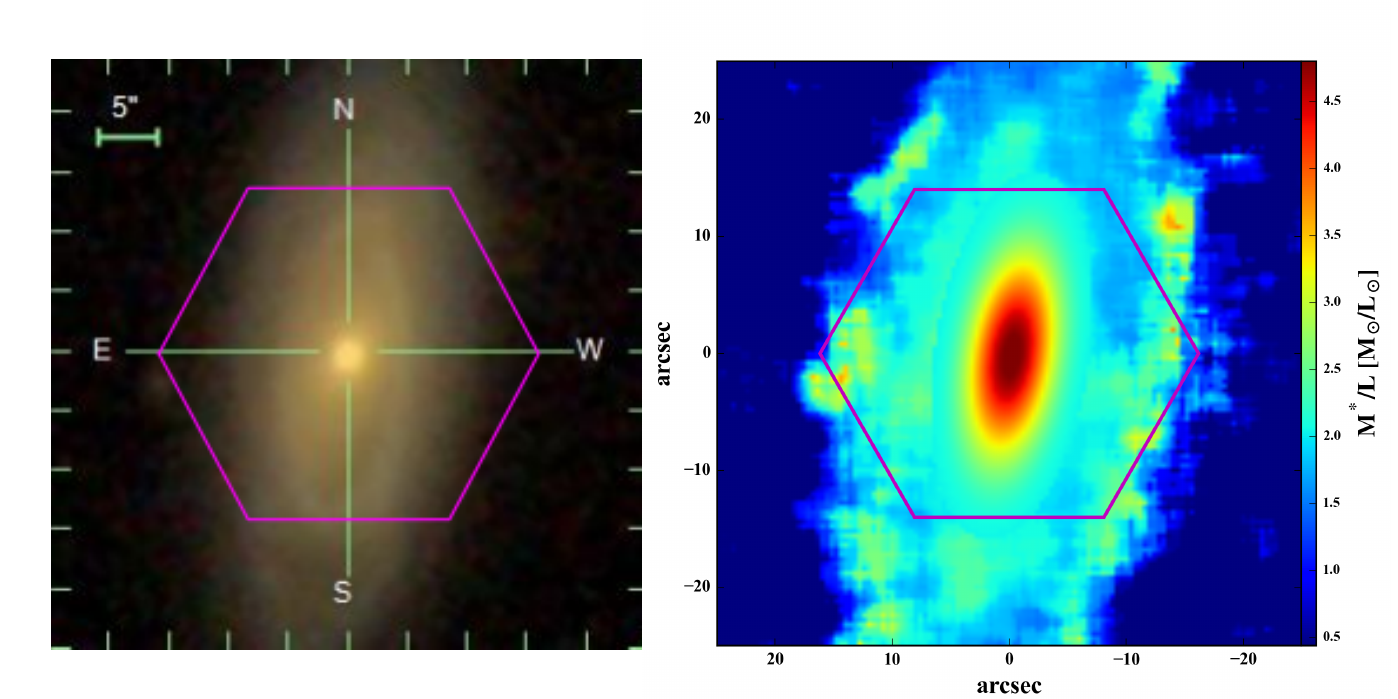}
\caption{The galaxy's SDSS 3--color image (left) and stellar mass-to-light ratio map (right). The magenta hexagon shows the region where IFU data are available. The inner stellar mass-to-light ratios are from MaNGA spectra and the outer are estimated by color map.}
\label{gradient1}
\end{center}
\end{figure*}

In order to describe the variation of the IMF, we define the IMF mismatch parameter similar to \citet{Treu2010}
\begin{equation}
\alpha_{\rm IMF} \equiv {\rm (M^*/L)_{JAM}^{nogas}/(M^*/L)_{SPS}},
\end{equation}
where $\alpha_{\rm IMF}$ is the ratio of the $\rm M^*/L$ values obtained by JAM and SPS for a Salpeter IMF.  In Fig.~\ref{IMF}, we plot the mismatch
parameter $\alpha_{\rm IMF}$ vs. the velocity dispersion within an effective radius.  The left panels are for the results from {\bf \scriptsize STARLIGHT} + BC03, 
while the right panels are for the results from {\bf \scriptsize pPXF} + MILES. 
The velocity dispersion is defined as 
\begin{equation}
\sigma_{e} = \sqrt{\langle{v^2_{\rm los}(<\rm{R_e})\rangle}},
\end{equation}
with $v_{\rm los}^2 \equiv V^2 + \sigma^2$, where V and $\sigma$ are the mean velocity and dispersion of the Gaussian which best fits the line-of-sight velocity
distribution.  A parameter value of $\alpha_{\rm IMF} = 1$ means that JAM and SPS give the same estimate.  

As can be seen from the panels, $\alpha_{\rm IMF}$ changes systematically with velocity dispersion. Galaxies with higher velocity dispersions are 
consistent with a Salpeter IMF, while galaxies with lower velocity dispersions are consistent with a Chabrier IMF. This is true for both elliptical and 
spiral galaxies, although there are larger scatters for spiral galaxies due to the effects from cold gas, dust extinction and larger degeneracies between
dark matter and stellar mass. 

We compare the systematic variation between different SPS software packages and templates in the left and right panels
 (left for {\bf \scriptsize STARLIGHT}+BC03, right for {\bf \scriptsize pPXF}+
MILES).  As can be seen, there are some small differences between the two approaches, however the trends are consistent within the statistical
errors as quantified in Table 1. The small differences can be understood as being due to the residual systematic differences between the two
 approaches, illustrated in Fig.~\ref{cmpML}.
 We fit the trend using a linear relation
 \begin{equation}
 \label{fitting}
\log \alpha_{\rm IMF} = a + b \times \log \sigma_{e}
\end{equation}
The fitting results are listed in Table~1 and plotted in Fig.~\ref{IMF}.

\begin{table}
\begin{center}
\label{IMF_fit}
\tabcolsep=2.0pt
\caption{The fitting coefficients of $\log \sigma_e$-$\log \alpha$ relation (Equation~\ref{fitting}) for elliptical and spiral galaxies}
\begin{tabular}{@{}lccc@{}}
\hline
\hline
 & $a$ & $b$ & $\Delta_{\rm int}$ \\
\hline
MaNGA elliptical ({\bf \scriptsize STARLIGHT}) & $-1.086\pm 0.006$ & $0.457\pm 0.033$ & 0.082\\
MaNGA elliptical ({\bf \scriptsize pPXF}) & $-1.399\pm 0.005$ & $0.591\pm 0.030$ & 0.063\\
MaNGA spiral ({\bf \scriptsize STARLIGHT})& $-1.364\pm 0.011$ & $0.596\pm 0.069$ & 0.156\\
MaNGA spiral ({\bf \scriptsize pPXF})& $-1.506\pm 0.011$ & $0.638\pm 0.075$ & 0.173\\
$\rm ATLAS^{3D}$ & $-0.895\pm 0.009$ & $0.364\pm 0.042$ & 0.083\\
\hline
\end{tabular}
\end{center}
\begin{minipage}{8.4cm}
Notes:
The units of $\log \sigma_{e}$ in the fitting are km/s. $\Delta_{\rm int}$ is the intrinsic scatter. The fitting is performed using the
{\bf \scriptsize lts\_linefit} software from \citet{Cappellari2013b}. 
\end{minipage}
\end{table}

The scatter for spiral galaxies with lower velocity dispersions is large since they are more affected by cold  gas and  dust extinction. 
The degeneracy between dark matter and stellar mass for these galaxies is also larger.  There are some outliers with low velocity 
dispersion but high $\alpha_{\rm IMF}$. We check the JAM and SPS model fitting results for these
galaxies, and find that they are galaxies with poor data quality (large uncertainties in dynamical modelling and SPS) or high inclination
(edge-on galaxies, strong dust extinction).
Our results are also consistent with results from gravitational lensing, albeit at slightly higher 
redshift $z=0.2$ \citep{Treu2010}. However, there are also a few discordant cases in \citet{Smith2013} and  \citet{Smith2015}.
In order to demonstrate that the IMF trend is not caused by the poor data quality of some galaxies, we plot these trends using the class A
subsample only in the lower panels of Fig.~\ref{IMF},  which have the most reliable fitting among the whole sample.

In addition to $\sigma_e$ (cf. eq.~5), which approximates the true projected velocity second moment, and includes contribution from both ordered rotation V 
and random motion $\sigma$, we also check whether $\alpha_{\rm IMF}$ depends on $\sigma$ alone, with the ordered rotation contribution removed. 
We additionally check the dependence against the specific stellar angular momentum 
parameter $\lambda_{\rm R_e}$ \citep{Emsellem2007} and metallicity [Z/H]. 
$\sigma$ is defined as 
\begin{equation}
\sigma = \sqrt{\langle{\sigma_{\rm los}^2(<\rm{R_e})\rangle}},
\end{equation}
which removes the velocity term in $\sigma_e$.
The results are shown in Fig.~\ref{metal}. The IMF trend is similar after changing $\sigma_e$ to $\sigma$.  For  $\lambda_{\rm R_e}$ and metallicity [Z/H],
there is no significant correlation with $\alpha_{\rm IMF}$. This is similar to the results from \citet{McDermid2014}, who found no strong correlation between IMF and 
metallicity using the same JAM modeling method.
\citet{Martin2015}, however, using line indices, showed that there is a correlation. This could be due to differences between the two methods.

More results about dark matter fractions, and the fundamental plane, stellar mass plane and mass plane 
scaling relationships will be given in a following paper (Li et al., in preparation).

\subsection{Effects of the stellar mass-to-light ratio gradient}\label{gradient}
\begin{figure}
\includegraphics[width=\columnwidth]{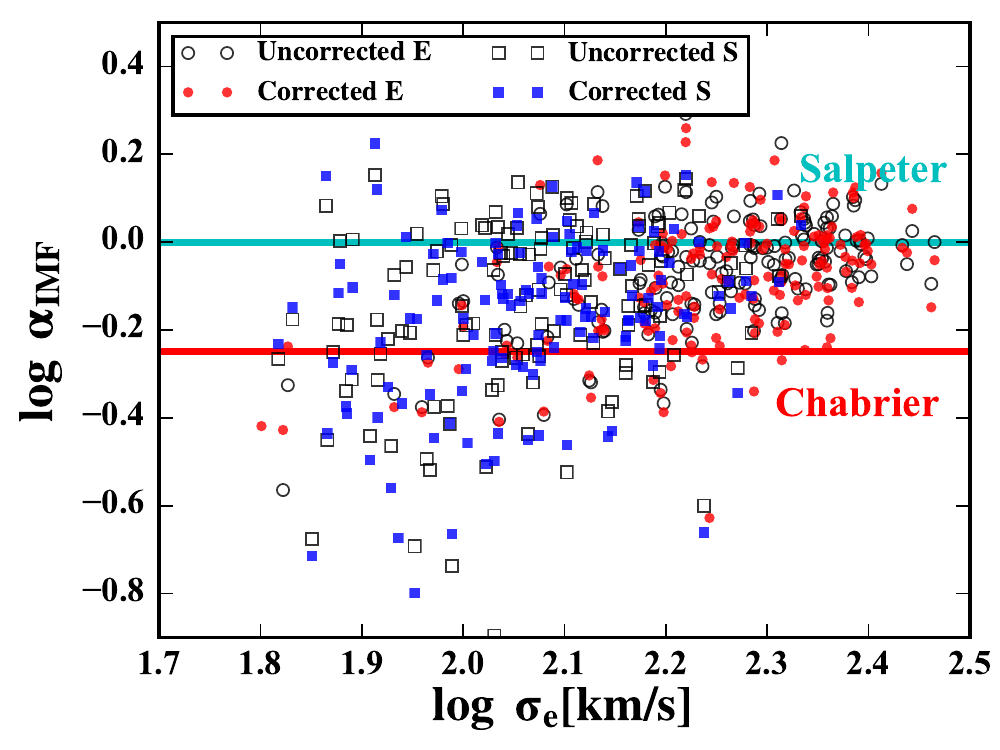}
\caption{
$\log \alpha_{\rm IMF}$ vs. $\log \sigma_e$ for class A galaxies with the stellar mass-to-light ratio gradient correction applied. 
The black void circles and squares 
show the positions of the galaxies without correction for elliptical and spiral galaxies respectively, the red circles and blue squares show the 
positions after correction. Other labels and legends are the same as Fig.~\ref{IMF}.}
\label{gradient2}
\end{figure}

For simplicity, when constructing mass models in JAM, a constant stellar mass-to-light ratio is usually assumed in order to convert a 
luminosity distribution to a stellar mass distribution, e.g. in \citet{Cappellari2013b}. Only recently have dynamical models started to include
spatial variations in the stellar mass-to-light ratio, in addition to allowing for a dark matter component \citep{Mitzkus2017, Poci2016}. This can be important, as a galaxy's stellar population may not be spatially uniform, so there may be a stellar mass-to-light ratio 
gradient, especially for younger galaxies (\citealt{Portinari2010,Tortora2011}; Ge et al., in preparation). In addition, different dust extinction
levels at different radii may also affect the mass-to-light ratio profile. It is important therefore to examine the effects caused by mass-to-light 
ratio gradients. 

In order to test the effects of such a gradient, we use the stellar mass profile directly in our mass models instead of the luminosity profile. 
In doing so, we avoid the assumption of a constant stellar mass-to-light ratio. The stellar mass profile is determined using our full spectrum fitting approach
 from the MaNGA spectra as described in Section 2.2. The spectra and kinematics are, by design, available over the same region. 
 This is the spatial region where the models are fitted and consequently is the region for which we can constrain the density profiles. 
 The results of dynamical models are only weakly dependent on the adopted stellar mass profiles at larger radii \citep{Krajnovic2005}.
 For this reason, the determination of accurate stellar mass profiles only really matters within the MaNGA field of view. 
 However, to avoid an abrupt and unphysical discontinuity in the stellar mass profile, outside the edge of the MaNGA field of view, 
 we use the more approximate color-$\rm M^*/L$ relation to smoothly extend the profile out to larger radii. 
 
In practice, to estimate the stellar mass density for our models, we start from the r-band image and multiply the surface brightness of each pixel contained within 
the MaNGA field of view by the stellar mass-to-light ratios measured from spectral fitting to obtain a stellar mass surface density map. At larger radii we estimate
the stellar mass-to-light ratios from galaxy's color. We take the SDSS g band and i band images \citep{Gunn2006,Eisenstein2011} and calculate the $g-i$ color in
each pixel. We apply the color-$\rm M^*/L$ relationship from \citet{Bell2003} to convert the color to the stellar mass-to-light ratio. We assume a Salpeter IMF in
the conversion. We use a median filter with window size 9 by 9 pixels (empirically chosen) to obtain a smoothed map. We scale the normalisation of the outer part
to match the  $\rm (M^*/L)_{SPS}$ values of the inner part around 1Re. This is to avoid a discontinuity in the profile,  although the scale factor is near 1 for majority
of the galaxies.  After obtaining the stellar mass surface density map, we perform the MGE fitting to it to obtain a stellar mass MGE. We then use this stellar mass
 MGE in our mass model  (the luminous MGE is still used as tracer density in JAM modeling). Since we already use stellar mass distribution in the mass model,
 the scale factor parameter in JAM is not $\rm M^*/L$ any more, but the mismatch parameter $\alpha_{\rm IMF}$ instead.

\begin{figure}
\includegraphics[width=\columnwidth]{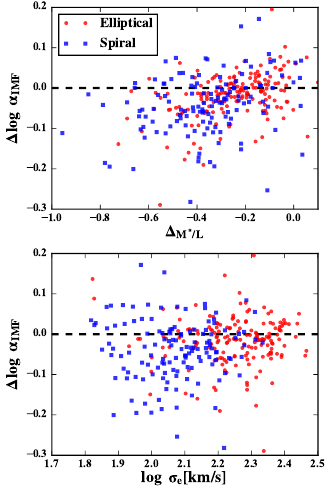}
\caption{
Change of the mismatch parameter ($\Delta \log \alpha_{\rm IMF}=\log \alpha_{\rm IMF}^{\rm corrected} -\log \alpha_{\rm IMF}^{\rm uncorrected}$) 
vs. gradient (upper) and velocity dispersion (lower). The red circles are elliptical galaxies and blue squares are spiral galaxies.
}
\label{gradient3}
\end{figure}

In Fig.~\ref{gradient1}, we show one galaxy as an example (MaNGA plate--IFU design : 8313--12705) which has one of the largest stellar 
mass-to-light ratio gradients in our galaxy sample.  As can be seen from the color map in the right panel of Fig.~\ref{gradient1}, the stellar 
mass-to-light ratio at the galaxy center is $\sim 5.0$, decreasing to $\sim 2.0$ in the outer regions.

We take Class A subset of galaxies and rerun JAM modeling with the gradient correction applied. 
In Fig.~\ref{gradient2}, we plot the IMF mismatch parameter $\alpha_{\rm IMF}$ vs. $\log \sigma_e$ as in Fig.~\ref{IMF} for these 
galaxies. As can be seen in the figure, even though there is some scatter after applying the gradient correction, the systematic 
variation still exists, and in fact it becomes even stronger.

In Fig.~\ref{gradient3}, we plot the change of mismatch parameter ($\Delta \log \alpha_{\rm IMF}=\log \alpha_{\rm IMF}^{\rm corrected}
 -\log \alpha_{\rm IMF}^{\rm uncorrected}$) vs. $\Delta_{\rm M^*/L}$ and $\log \sigma_e$. $\Delta_{\rm M^*/L}$ is the stellar 
 mass-to-light ratio gradient obtained by fitting the linear function $\log {\rm M^*/L} = a+ b\log R$ to the MaNGA IFU results.  
 As can be seen in the top panel, the change 
 in the mismatch parameter increases as the gradient becomes larger (more negative). When the gradient is close to 0, the 
 $\alpha_{\rm IMF}$ before and after correction has no systematic difference. However when the gradient increases,  $\alpha_{\rm IMF}$
 systematically decreases after gradient correction (implying lighter IMFs). In the lower panel, one can see that galaxies with lower velocity 
 dispersions have systematically smaller $\alpha_{\rm IMF}$ after the gradient correction (this means the systematic IMF trend become 
 stronger after correcting for the gradient effect). This is because lower dispersion galaxies have younger stellar populations and steeper 
 gradients. Our test suggests that the IMF trend is even stronger than what we determined for the case without gradients 
 as well as in previously published studies.

\subsection{Effect of inclination and dust extinction for spiral galaxy}\label{incAndextinct}
In this section, we discuss the effects of inclination and dust extinction on spiral galaxies. Since observations suffer more from dust
extinction when galaxies are nearly edge-on, the dynamical M/L of flat galaxies tends to be overestimated when galaxies are 
nearly face-on \citep{Lablanche2012}. 

In the top and middle panels of Fig.~\ref{inclination}, we show 
$\log \alpha_{\rm IMF}$ vs. $\log \sigma_e$ for all the spiral galaxies in our sample with different observed axis ratios (i.e. inclinations) 
and dust extinction. As can be seen, edge-on galaxies or galaxies with higher dust extinction
are slightly biased to higher  $\alpha_{\rm IMF}$. This may be because SPS underestimates the dust extinction or the age of these
galaxies, which leads to a lower  $\rm (M^*/L)_{SPS}$.  For intermediate and low inclination/dust extinction galaxies, there
is no systematic difference.  In the bottom panel of Fig.~\ref{inclination}, we show the value of extinction predicted by our 
{\bf \scriptsize STARLIGHT} SPS fits
vs.  axis ratio (i.e. inclination). As expected, edge-on galaxies have more dust extinction, and our trend is similar to the results using 
optical and near infrared photometry data obtained by \citet{Devour2016}. In Fig~\ref{cmpExtinct}, we further compare the $\rm M^*/L_{SPS}$ assuming
a CAL extinction law in SPS with the $\rm M^*/L_{SPS}$ assuming a CCM \citep{Cardelli1989} extinction law and find no significant difference. 

\begin{figure}
\includegraphics[width=\columnwidth]{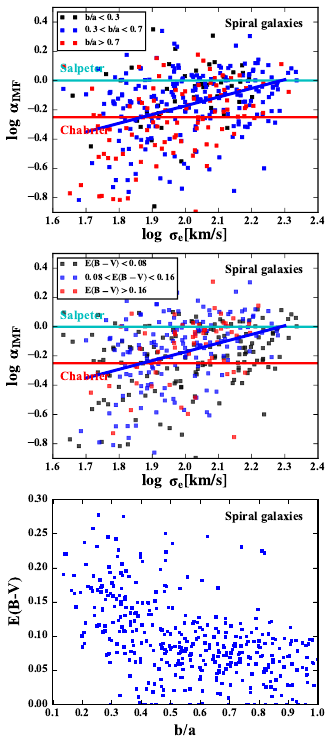}
\caption{
Top: $\log \alpha_{\rm IMF}$ vs. $\log \sigma_e$ for all the spiral galaxies with different observed axis ratios. The black squares are 
for near edge-on sample ($b/a < 0.3$), blue squares for intermediate inclined sample ($0.3<b/a < 0.7$) and red squares for near 
face-on sample ($b/a>0.7$). The blue solid line shows the fitting results for spiral galaxies in Table~1.
 Middle:
 $\log \alpha_{\rm IMF}$ vs. $\log \sigma_e$ for all the spiral galaxies with different extinction values predicted by SPS. 
 The black squares are for low extinction sample ($\rm E(B-V) < 0.08$), blue squares for intermediate extinction sample ($\rm 0.08<E(B-V) < 0.16$) 
 and red squares for high extinction sample ($\rm E(B-V) > 0.16$). The blue solid line shows the fitting results for spiral galaxies in Table~1.
 Bottom: dust extinction values predicted by SPS ({\bf \scriptsize STARLIGHT}) vs. observed axis
 ratios.  Other labels are the same as Fig.~\ref{IMF}.
}
\label{inclination}
\end{figure}

\begin{figure}
\includegraphics[width=\columnwidth]{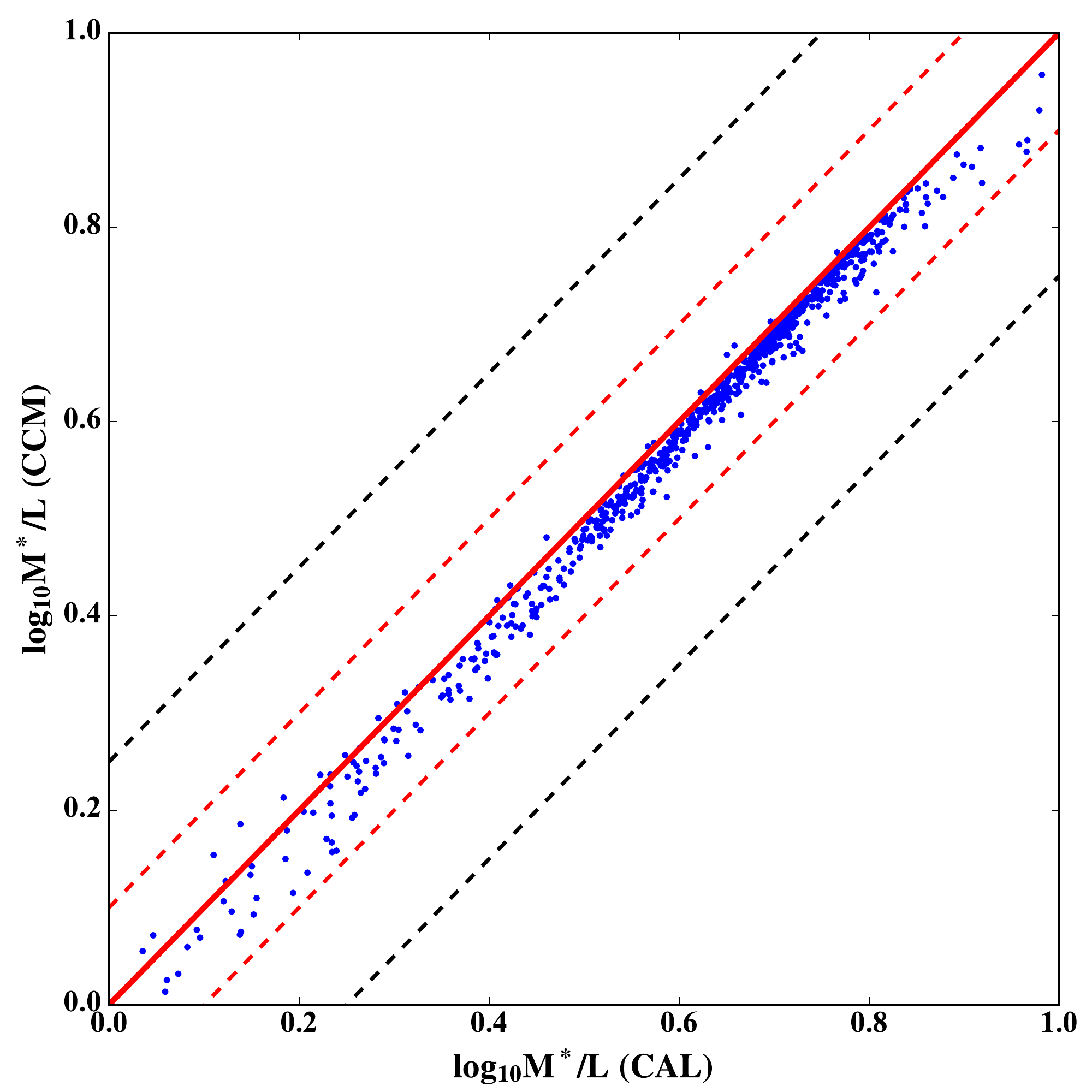}
\caption{
Comparison between the SPS stellar mass-to-light ratios determined using both the CAL and CCM extinction laws.  The lines are the same as Fig.~\ref{cmpML}.
}
\label{cmpExtinct}
\end{figure}

\section{Conclusions}\label{conclusion}
We have performed JAM modeling for 816 galaxies, with good data quality, from the MaNGA DR13 sample, including both elliptical 
and spiral galaxies. We have compared the stellar mass-to-light ratios from SPS and JAM modeling, and find a systematic variation 
of the initial mass function for both elliptical and spiral galaxies. Galaxies with lower velocity dispersions within an effective radius 
are consistent with a Chabrier-like IMF,
while galaxies with higher velocity dispersions are consistent with a more bottom heavy IMF like the Salpeter IMF. These results agree
 well with  previous studies (e.g. \citealt{Conroy2012, Dutton2013, Cappellari2012, Cappellari2013a, Posacki2015}).  

In previous IMF studies using stellar dynamics or gravitational lensing, a constant stellar mass-to-light ratio was assumed. 
However, there are stellar mass-to-light ratio
gradients especially for young galaxies. So, in addition, we have examined the effect of this gradient.  
We use our Class A galaxies to introduce this gradient and performed a comparison test. We found that the systematic 
IMF trend still exists, and becomes even stronger after the gradient correction. In addition to the stellar mass-to-light ratio gradient,
there are also studies which showed that the IMF inside a galaxy could also be different \citep{Martin2015, Barbera2016}. In their studies,
they found that for several early type galaxies, the IMF is bottom heavy in the central region, but bottom light in the outer region. This will
lead to an even steeper stellar mass-to-light ratio gradient and have some effects on our results. However, the systematic IMF trend
in this work is based on a globally averaged IMF for a galaxy. The IMF variation inside a galaxy is beyond  the 
scope of this paper, but we will return to issue in a future work.

Spiral galaxies with lower velocity dispersions have results with large scatters. This is because
they are more affected by cold gas and dust extinction. The degeneracy between dark matter and stellar mass is also stronger in these 
galaxies. Our results show that these galaxies favour a Chabrier-like IMF, and this is consistent with 
the results in \citet{Bershady2011,Dutton2011,Brewer2012}. Galaxy inclinations do not have strong effects except for nearly 
edge-on galaxies (higher dust extinction leads to larger uncertainty in SPS).

Observationally it will be interesting to examine further whether there are differences in the IMF between galaxy discs and bulges in spiral galaxies (e.g.
 \citealt{Dutton2013b}). Theoretically it is unclear how the IMF changes when two galaxies with different IMFs merge, and whether the IMF variation 
changes as a function of redshift. If it does, how this changes the stellar mass function of galaxies and the evolution of the stellar mass as a function of 
cosmic time needs investigation.
 
\section*{Acknowledgements}
HL would like to thank Drs. Yiping Shu and Zheng Zheng for many useful discussions,
and Drs. Anne Jaskot and David Stark for advice on the gas fraction for spiral galaxies.
MC acknowledges support from a Royal Society University Research Fellowship.
We performed our computer runs on the Zen high performance
computer cluster of the National Astronomical Observatories,
 Chinese Academy of Sciences (NAOC). This work has
been supported by the Strategic Priority Research Program
``The Emergence of Cosmological Structures" of the Chinese 
Academy of Sciences Grant No. XDB09000000 (RJL
and SM), and by the National Natural Science Foundation of
China (NSFC) under grant numbers 11333003, 11390372
(SM), and 11303033, 11511130054, 11333001 (RL), by the Newton Fund,
 by the Youth Innovation Promotion Association of CAS, and by
World Premier International Research Center Initiative (WPI Initiative), MEXT, Japan.  

Funding for the Sloan Digital Sky Survey IV has been provided by
the Alfred P. Sloan Foundation, the U.S. Department of Energy Office of
Science, and the Participating Institutions. SDSS-IV acknowledges
support and resources from the Center for High-Performance Computing at
the University of Utah. The SDSS web site is www.sdss.org.

SDSS-IV is managed by the Astrophysical Research Consortium for the 
Participating Institutions of the SDSS Collaboration including the 
Brazilian Participation Group, the Carnegie Institution for Science, 
Carnegie Mellon University, the Chilean Participation Group, the French Participation Group, 
Harvard-Smithsonian Center for Astrophysics, 
Instituto de Astrof\'isica de Canarias, The Johns Hopkins University, 
Kavli Institute for the Physics and Mathematics of the Universe (IPMU) / 
University of Tokyo, Lawrence Berkeley National Laboratory, 
Leibniz Institut f\"ur Astrophysik Potsdam (AIP),  
Max-Planck-Institut f\"ur Astronomie (MPIA Heidelberg), 
Max-Planck-Institut f\"ur Astrophysik (MPA Garching), 
Max-Planck-Institut f\"ur Extraterrestrische Physik (MPE), 
National Astronomical Observatories of China, New Mexico State University, 
New York University, University of Notre Dame, 
Observat\'ario Nacional / MCTI, The Ohio State University, 
Pennsylvania State University, Shanghai Astronomical Observatory, 
United Kingdom Participation Group,
Universidad Nacional Aut\'onoma de M\'exico, University of Arizona, 
University of Colorado Boulder, University of Oxford, University of Portsmouth, 
University of Utah, University of Virginia, University of Washington, University of Wisconsin, 
Vanderbilt University, and Yale University.





\begin{thebibliography}{99}
\bibitem[\protect\citeauthoryear{Barnab{\`e} et al.}{2012}]{Barnab2012} Barnab{\`e} M., et al., 2012, MNRAS, 423, 1073
\bibitem[\protect\citeauthoryear{Bastian, Covey, \& Meyer}{2010}]{Bastian2010} Bastian N., Covey K.~R., Meyer M.~R., 2010, ARA\&A, 48, 339
\bibitem[\protect\citeauthoryear{Bell et al.}{2003}]{Bell2003} Bell E.~F., McIntosh D.~H., Katz N., Weinberg M.~D., 2003, ApJS, 149, 289
\bibitem[\protect\citeauthoryear{Bershady et al.}{2011}]{Bershady2011} Bershady M.~A., Martinsson T.~P.~K., Verheijen M.~A.~W., Westfall K.~B., Andersen D.~R., Swaters R.~A., 2011, ApJ, 739, L47 
\bibitem[\protect\citeauthoryear{Bigiel \& Blitz}{2012}]{Bigiel2012} Bigiel F., Blitz L., 2012, ApJ, 756, 183 
\bibitem[\protect\citeauthoryear{Brewer et al.}{2012}]{Brewer2012} Brewer B.~J., et al., 2012, MNRAS, 422, 3574 
\bibitem[\protect\citeauthoryear{Bruzual \& Charlot}{2003}]{BC03} Bruzual G., Charlot S., 2003, MNRAS, 344, 1000
\bibitem[\protect\citeauthoryear{Bryant et al.}{2015}]{Bryant2015} Bryant J.~J., et al., 2015, MNRAS, 447, 2857
\bibitem[\protect\citeauthoryear{Bundy et al.}{2015}]{Bundy2015} Bundy K., et al., 2015, ApJ, 798, 7 
\bibitem[\protect\citeauthoryear{Calzetti et al.}{2000}]{Calzetti2000} Calzetti D., Armus L., Bohlin R.~C., Kinney A.~L., Koornneef J., Storchi-Bergmann T., 2000, ApJ, 533, 682 
\bibitem[\protect\citeauthoryear{Cappellari}{2002}]{Cappellari2002} Cappellari M., 2002, MNRAS, 333, 400
\bibitem[\protect\citeauthoryear{Cappellari \& Copin}{2003}]{Cappellari2003} Cappellari M., Copin Y., 2003, MNRAS, 342, 345
\bibitem[\protect\citeauthoryear{Cappellari \& Emsellem}{2004}]{Cappellari2004} Cappellari M., Emsellem E., 2004, PASP, 116, 138
\bibitem[\protect\citeauthoryear{Cappellari et al.}{2006}]{Cappellari2006} Cappellari M., et al., 2006, MNRAS, 366, 1126 
\bibitem[\protect\citeauthoryear{Cappellari}{2008}]{Cappellari2008} Cappellari M., 2008, MNRAS, 390, 71
\bibitem[\protect\citeauthoryear{Cappellari et al.}{2012}]{Cappellari2012} Cappellari M., et al., 2012, Natur, 484, 485 
\bibitem[\protect\citeauthoryear{Cappellari et al.}{2013a}]{Cappellari2013a} Cappellari M., et al., 2013a, MNRAS, 432, 1862
\bibitem[\protect\citeauthoryear{Cappellari et al.}{2013b}]{Cappellari2013b} Cappellari M., et al., 2013b, MNRAS, 432, 1709
\bibitem[\protect\citeauthoryear{Cappellari}{2016}]{Cappellari2016} Cappellari M., 2016, ARA\&A, 54, 597
\bibitem[\protect\citeauthoryear{Cappellari}{2017}]{Cappellari2017} Cappellari M., 2017, MNRAS, 466, 798
\bibitem[\protect\citeauthoryear{Cardelli, Clayton, \& Mathis}{1989}]{Cardelli1989} Cardelli J.~A., Clayton G.~C., Mathis J.~S., 1989, ApJ, 345, 245 
\bibitem[\protect\citeauthoryear{Chabrier}{2003}]{Chabrier2003} Chabrier G., 2003, PASP, 115, 763
\bibitem[\protect\citeauthoryear{Cid Fernandes et al.}{2005}]{Fernandes2005} Cid Fernandes R., Mateus A., Sodr{\'e} L., Stasi{\'n}ska G., Gomes J.~M., 2005, MNRAS, 358, 363 
\bibitem[\protect\citeauthoryear{Combes et al.}{2013}]{Combes2013} Combes F., Garc{\'{\i}}a-Burillo S., Braine J., Schinnerer E., Walter F., Colina L., 2013, A\&A, 550, A41 
\bibitem[\protect\citeauthoryear{Conroy \& van Dokkum}{2012}]{Conroy2012} Conroy C., van Dokkum P.~G., 2012, ApJ, 760, 71 
\bibitem[\protect\citeauthoryear{Devour \& Bell}{2016}]{Devour2016} Devour B.~M., Bell E.~F., 2016, MNRAS, 459, 2054 
\bibitem[\protect\citeauthoryear{Drory et al.}{2015}]{Drory2015} Drory N., et al., 2015, AJ, 149, 77
\bibitem[\protect\citeauthoryear{Dutton et al.}{2011}]{Dutton2011} Dutton A.~A., et al., 2011, MNRAS, 416, 322
\bibitem[\protect\citeauthoryear{Dutton et al.}{2013a}]{Dutton2013} Dutton A.~A., Macci{\`o} A.~V., Mendel J.~T., Simard L., 2013, MNRAS, 432, 2496 
\bibitem[\protect\citeauthoryear{Dutton et al.}{2013b}]{Dutton2013b} Dutton A.~A., et al., 2013, MNRAS, 428, 3183 
\bibitem[\protect\citeauthoryear{Dutton et al.}{2015}]{Dutton2015} Dutton A.~A., Macci{\`o} A.~V., Stinson G.~S., Gutcke T.~A., Penzo C., Buck T., 2015, MNRAS, 453, 2447 

\bibitem[\protect\citeauthoryear{Eisenstein et al.}{2011}]{Eisenstein2011} Eisenstein D.~J., et al., 2011, AJ, 142, 72 
\bibitem[\protect\citeauthoryear{Emsellem, Monnet, \& Bacon}{1994}]{Emsellem1994} Emsellem E., Monnet G., Bacon R., 1994, A\&A, 285, 723
\bibitem[\protect\citeauthoryear{Emsellem et al.}{2007}]{Emsellem2007} Emsellem E., et al., 2007, MNRAS, 379, 401 
\bibitem[\protect\citeauthoryear{Foreman-Mackey et al.}{2013}]{Foreman2013} Foreman-Mackey D., Hogg D.~W., Lang D., Goodman J., 2013, PASP, 125, 306 
\bibitem[\protect\citeauthoryear{Gunn et al.}{2006}]{Gunn2006} Gunn J.~E., et al., 2006, AJ, 131, 2332
\bibitem[\protect\citeauthoryear{Hogg et al.}{2002}]{Hogg2002} Hogg D.~W., Baldry I.~K., Blanton M.~R., Eisenstein D.~J., 2002, astro, arXiv:astro-ph/0210394 
\bibitem[\protect\citeauthoryear{Huang et al.}{2012}]{Huang2012} Huang S., Haynes M.~P., Giovanelli R., Brinchmann J., 2012, ApJ, 756, 113
\bibitem[\protect\citeauthoryear{Jaskot et al.}{2015}]{Jaskot2015} Jaskot A.~E., Oey M.~S., Salzer J.~J., Van Sistine A., Bell E.~F., Haynes M.~P., 2015, ApJ, 808, 66 
\bibitem[\protect\citeauthoryear{Krajnovi{\'c} et al.}{2005}]{Krajnovic2005} Krajnovi{\'c} D., Cappellari M., Emsellem E., McDermid R.~M., de Zeeuw P.~T., 2005, MNRAS, 357, 1113 
\bibitem[\protect\citeauthoryear{La Barbera et al.}{2016}]{Barbera2016} La Barbera F., Vazdekis A., Ferreras I., Pasquali A., Cappellari M., Mart{\'{\i}}n-Navarro I., Sch{\"o}nebeck F., Falc{\'o}n-Barroso J., 2016, MNRAS, 457, 1468 

\bibitem[\protect\citeauthoryear{Lablanche et al.}{2012}]{Lablanche2012} Lablanche P.-Y., et al., 2012, MNRAS, 424, 1495
\bibitem[\protect\citeauthoryear{Law et al.}{2015}]{Law2015} Law D.~R., et al., 2015, AJ, 150, 19   
\bibitem[\protect\citeauthoryear{Law et al.}{2016}]{Law2016} Law D.~R., et al., 2016, AJ, 152, 83
\bibitem[\protect\citeauthoryear{Li et al.}{2016}]{Hongyu2016} Li H., Li R., Mao S., Xu D., Long R.~J., Emsellem E., 2016, MNRAS, 455, 3680
\bibitem[\protect\citeauthoryear{Lintott et al.}{2008}]{Lintott2008} Lintott C.~J., et al., 2008, MNRAS, 389, 1179 
\bibitem[\protect\citeauthoryear{Mart{\'{\i}}n-Navarro et al.}{2015}]{Martin2015} Mart{\'{\i}}n-Navarro I., Barbera F.~L., Vazdekis A., Falc{\'o}n-Barroso J., Ferreras I., 2015, MNRAS, 447, 1033 
\bibitem[\protect\citeauthoryear{McDermid et al.}{2014}]{McDermid2014} McDermid R.~M., et al., 2014, ApJ, 792, L37 
\bibitem[\protect\citeauthoryear{Mitzkus, Cappellari, \& Walcher}{2017}]{Mitzkus2017} Mitzkus M., Cappellari M., Walcher C.~J., 2017, MNRAS, 464, 4789 
\bibitem[\protect\citeauthoryear{Panter et al.}{2007}]{Panter2007} Panter B., Jimenez R., Heavens A.~F., Charlot S., 2007, MNRAS, 378, 1550
\bibitem[\protect\citeauthoryear{Poci, Cappellari, \& McDermid}{2016}]{Poci2016} Poci A., Cappellari M., McDermid R.~M., 2016, arXiv, arXiv:1612.05805
\bibitem[\protect\citeauthoryear{Posacki et al.}{2015}]{Posacki2015} Posacki S., Cappellari M., Treu T., Pellegrini S., Ciotti L., 2015, MNRAS, 446, 493 
\bibitem[\protect\citeauthoryear{Portinari \& Salucci}{2010}]{Portinari2010} Portinari L., Salucci P., 2010, A\&A, 521, A82
\bibitem[\protect\citeauthoryear{Salpeter}{1955}]{Salpeter1955} Salpeter E.~E., 1955, ApJ, 121, 161 
\bibitem[\protect\citeauthoryear{S{\'a}nchez-Bl{\'a}zquez et al.}{2006}]{MILES2006} S{\'a}nchez-Bl{\'a}zquez P., et al., 2006, MNRAS, 371, 703 
\bibitem[\protect\citeauthoryear{Schaller et al.}{2015}]{Schaller2015} Schaller M., et al., 2015, MNRAS, 451, 1247 

\bibitem[\protect\citeauthoryear{Spiniello et al.}{2012}]{Spiniello2012} Spiniello C., Trager S.~C., Koopmans L.~V.~E., Chen Y.~P., 2012, ApJ, 753, L32

\bibitem[\protect\citeauthoryear{S{\'a}nchez et al.}{2012}]{Sanchez2012} S{\'a}nchez S.~F., et al., 2012, A\&A, 538, A8
\bibitem[\protect\citeauthoryear{SDSS Collaboration et al.}{2016}]{DR13} SDSS Collaboration, et al., 2016, arXiv, arXiv:1608.02013 
\bibitem[\protect\citeauthoryear{Smee et al.}{2013}]{Smee2013} Smee S.~A., et al., 2013, AJ, 146, 32 
\bibitem[\protect\citeauthoryear{Smith \& Lucey}{2013}]{Smith2013} Smith R.~J., Lucey J.~R., 2013, MNRAS, 434, 1964 
\bibitem[\protect\citeauthoryear{Smith et al.}{2015}]{Smith2015} Smith R.~J., Alton P., Lucey J.~R., Conroy C., Carter D., 2015, MNRAS, 454, L71 
\bibitem[\protect\citeauthoryear{Tortora et al.}{2009}]{Tortora2009} Tortora C., Napolitano N.~R., Romanowsky A.~J., Capaccioli M., Covone G., 2009, MNRAS, 396, 1132
\bibitem[\protect\citeauthoryear{Tortora et al.}{2011}]{Tortora2011} Tortora C., Napolitano N.~R., Romanowsky A.~J., Jetzer P., Cardone V.~F., Capaccioli M., 2011, MNRAS, 418, 1557 
\bibitem[\protect\citeauthoryear{Tollet et al.}{2016}]{Tollet2016} Tollet E., et al., 2016, MNRAS, 456, 3542 
\bibitem[\protect\citeauthoryear{Treu et al.}{2010}]{Treu2010} Treu T., Auger M.~W., Koopmans L.~V.~E., Gavazzi R., Marshall P.~J., Bolton A.~S., 2010, ApJ, 709, 1195 
\bibitem[\protect\citeauthoryear{Vazdekis et al.}{2010}]{Vazdekis2010} Vazdekis A., S{\'a}nchez-Bl{\'a}zquez P., Falc{\'o}n-Barroso J., Cenarro A.~J., Beasley M.~A., Cardiel N., Gorgas J., Peletier R.~F., 2010, MNRAS, 404, 1639 
\bibitem[\protect\citeauthoryear{Yan et al.}{2016a}]{Yan2016a} Yan R., et al., 2016a, AJ, 151, 8  
\bibitem[\protect\citeauthoryear{Yan et al.}{2016b}]{Yan2016b} Yan R., et al., 2016b, AJ, 152, 197
\end{thebibliography}



\label{lastpage}
\end{document}